\documentclass[twocolumn,aps,prl,showpacs,floatfix, superscriptaddress]{revtex4}
\usepackage[pdftex,colorlinks,breaklinks,plainpages=false,citecolor=blue,linkcolor=blue,urlcolor
=blue,pdfstartview=FitH]{hyperref}
\usepackage[dvips]{graphicx}
\usepackage{amsmath}
\usepackage{color}
\usepackage{amsfonts}
\usepackage{amssymb}
\usepackage{epstopdf}
\DeclareGraphicsExtensions{.eps,.png,.pdf,.tiff}
\begin{document}

\title{Experimental observation of extreme multistability \\
in an electronic system of two coupled R\"{o}ssler oscillators}
\author{Mitesh S. Patel }
\author{Unnati Patel}
\author{Abhijit Sen\footnote{email:senabhijit@gmail.com}}
\author{Gautam C. Sethia}
\affiliation{Institute for Plasma Research, Bhat, Gandhinagar 382 428, India}
\author{Chittaranjan Hens}
\author{Syamal K. Dana}
\affiliation{CSIR-Indian Institute of Chemical Biology, Kolkata, India}
\author{Ulrike Feudel}
\affiliation{Institute for Chemistry and Biology of the Marine Environment, University of Oldenburg, Oldenberg, Germany and
Institute for Physical Science and Technology, University of Maryland, College Park, MD 20742-2431, U.S.A.}
\author{Kenneth Showalter}
\affiliation{ West Virginia University, Morgantown,
West Virginia 26506-6045, U.S.A.}
\author{Calistus N. Ngonghala}
\affiliation{National Institute for Mathematical and Biological Synthesis, UT Knoxville, TN 37996, USA}
\author{Ravindra E. Amritkar}
\affiliation{Physical Research Laboratory, Ahmedabad 380009, India}
\date{\today}


\begin{abstract}
We report the first experimental observation of  extreme multistability in a controlled laboratory
investigation. Extreme multistability arises when infinitely many attractors coexist for the same
set of system parameters. The behavior was predicted earlier on theoretical grounds, supported by
numerical studies of models of two coupled identical or nearly identical systems. We construct
and couple two analog circuits based on a modified coupled R\"{o}ssler system and demonstrate the occurrence of
extreme multistability through a controlled switching to different
attractor states purely through a change in initial conditions for a fixed set of system parameters.
Numerical studies of the coupled model equations are in agreement with our experimental
findings.
\end{abstract}



\maketitle
Multistability is a common occurrence in many nonlinear dynamical systems, corresponding to
the coexistence of more than one stable attractor for the same set of system parameters
\cite{feudel08}. A large number of theoretical and experimental studies have explored this
phenomenon in a variety of physical \cite{arecchi82,masoller02,kastrup94,schwarz00}, chemical
\cite{ganapathy84,marmillot91} and biological \cite{foss96,huisman01} systems. A curious and
novel manifestation of this phenomenon arises when a system can have an infinite number of
coexisting attractors, where each attractor is associated with a particular set of initial
conditions \cite{sun99,calistus11}.  Extreme multistability 
was first demonstrated in a system of two coupled identical
Lorenz oscillators by Sun {\it et al.} \cite{sun99} and subsequently investigated in the
three-variable autocatalator model by Ngonghala {\it et al.} \cite{calistus11}. More recently
Hens {\it et al.} \cite{hens12} have demonstrated the existence of extreme multistability  in a system of
two coupled R\"{o}ssler oscillators and in a chemical autocatalator model \cite{Peng1990}.

In all these studies, a special coupling was applied between two three-variable chaotic systems to
form six-variable coupled systems. Numerical simulations of the synchronization of the coupled systems were carried
out for a fixed set of system parameters and only the initial conditions were changed. The synchronized
systems were found to evolve to different attractor states (fixed points, limit cycles, chaotic
states) purely through changes in initial conditions, typically with a change in the initial condition
of just one of the state variables. The origin of this behaviour lies in the synchronization dynamics of the two coupled 
subsystems. Two properties are essential for the appearance of extreme multistability in two coupled $n$-dimensional 
nonlinear systems: (i) the complete synchronization of $n-1$ state variables of the two systems and (ii) the synchronization 
of the remaining state variable in each subsystem according to a conserved quantity $K$ in the long-term limit $t \to \infty$. 
The conserved quantity has a profound effect on the synchronization dynamics, it characterizes the synchronization manifold. It 
also leads to a neutrally stable direction 
in the steady states and orbits which gives rise to a dependence on the initial conditions of the asymptotic state at $t \to \infty$. 
In addition, perturbations give rise to new dynamical states, as the system is shifted from one synchronization manifold to another.

Extreme multistability might have important consequences in the reproducibility of certain
experimental systems. For example, some chemical reactions, such as the chlorite-thiosulfate
reaction \cite{orban82} and the chlorite-iodide reaction \cite{nagypal86} consistently exhibit
irreproducibility: despite great care to ensure reproducibility, these reactions show a random
long-term behavior for the same set of experimental conditions. The cause of the
irreproducibility is not known; extreme multistability offers a possible mechanism for
the behavior. 
However, to the
best of our knowledge, there is no direct experimental verification of this new type of
dynamical behavior in a controlled laboratory investigation. In this Letter, we report experimental
observations of a coupled electronic circuit system that displays extreme multistability.
\begin{figure}
\raisebox{5.4cm} {(a)}\includegraphics[width =8.5cm]{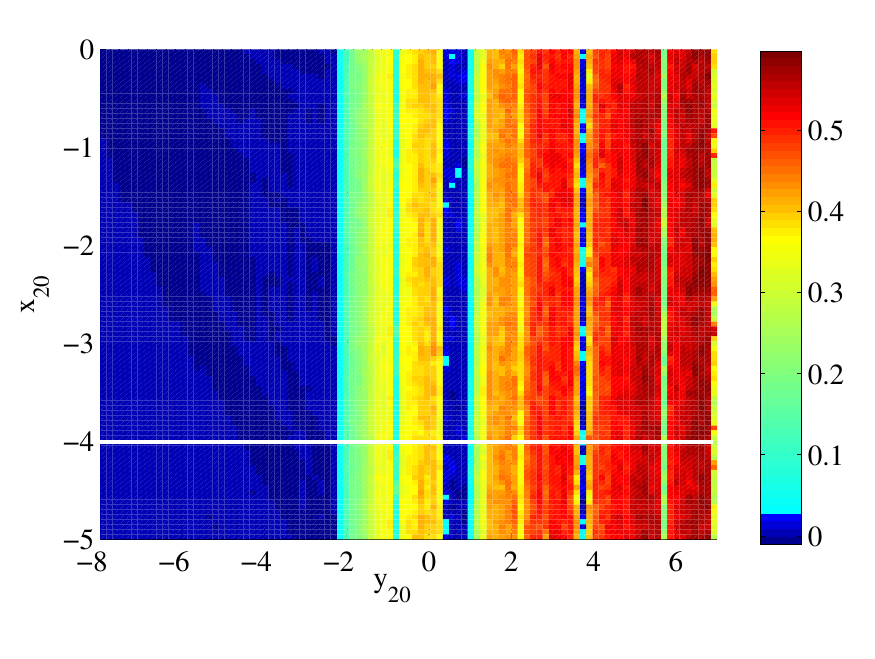}  \\
\raisebox{4.5cm} {(b)}\includegraphics[width = 8.0cm]{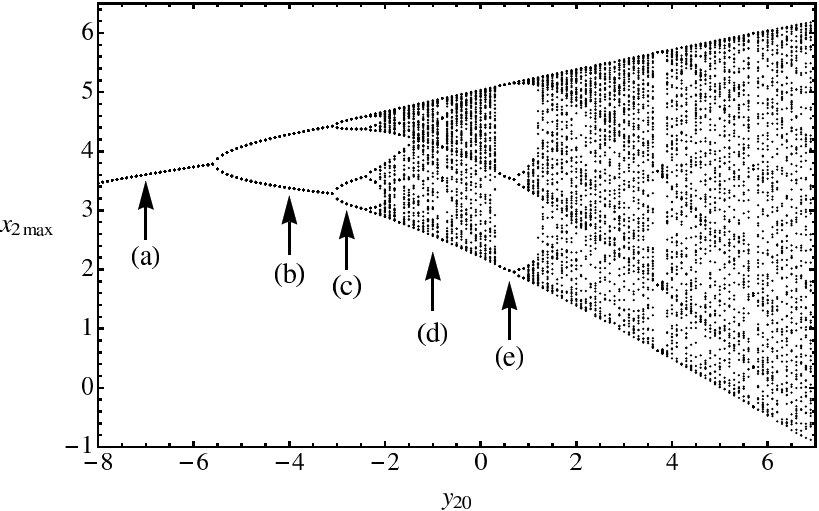}  
\caption{(a) Plot of the maximum Lyapunov exponent
(color coded) in the ($x_{20} , y_{20}$) space with parameter values fixed
at $\alpha= 0.02, a=0.2, b=0.2,c=5.7$ and the other initial conditions fixed at 
$x_{10}=y_{10}=z_{10}=z_{20} =0$,  (b) Maxima of $x_2$ plotted against initial values of $y_2$ 
along the line $x_{20}=-4.0$.} 
\label{fig:fig1}
\end{figure}

Our experiments are carried out on an analog circuit system closely based on the model
investigated by Hens {\it et al.} \cite{hens12}, consisting of a set of coupled R\"{o}ssler equations,
namely,
\begin{subequations}
\label{eqn}
\begin{align}
\dot{x}_1 & =   -y_1 - z_1 \label{a}\\
\dot{y}_1 & =    x_2 + a y_2 + \alpha (x_2-x_1) \label{b}\\
\dot{z}_1 & =    b +2 x_2 z_2 -c z_1 \label{c}\\
\dot{x}_2 & =    x_1 - x_2 -y_1 - z_1 \label{d}\\
\dot{y}_2 & =    x_2+a y_2 \label{e}\\
\dot{z}_2 & =    b +2 x_2 z_2 - c z_2 \label{f}
\end{align}
\end{subequations}
where $\alpha$, $a$, $b$ and $c$ are constants (with $c >0$) and $(x_1,y_1,z_1)$ and $(x_2,y_2, z_2)$ are the
state variables of the two subsystems. If we set $x_1=x_2$, $y_1=y_2$ and $z_1=z_2$ then the two subsystems become decoupled
and each individual subsystem represents a R\"{o}ssler oscillator. The factor of two multiplying the nonlinear terms in \eqref{c} and \eqref{f}
arises from a scaling down of the original R\"{o}ssler system variables by a factor of $2$. This is done to restrict the output signal voltage
range, in the circuit implementation of the equations, to within $\pm 15$ volts in order to avoid saturation of the circuit.  The coupled
system \eqref{eqn} is a variant of the set  analyzed in \cite{hens12} in that \eqref{b} of our system is different from the corresponding equation 
in \cite{hens12}. Our system becomes identical to that of \cite{hens12} for $\alpha=-1$. 
However the basic property of extreme multistability is still preserved in the modified system as can be seen from an analysis of a reduced set of 
equations that govern the differences (``errors'') of the corresponding state variables of the subsystems, namely, 
\begin{subequations}
\label{error}
\begin{align}
\dot{e}_1 & =   -e_1 \label{ea} \\ 
\dot{e}_2 & =    -\alpha e_{1} \label{eb} \\ 
\dot{e}_3 & =    -c e_3 \label{ec}
\end{align}
\end{subequations}
where $e_1=x_1-x_2$, $e_2=y_1-y_2$ and $e_3=z_1-z_2$. Upon complete synchronization, the error dynamics evolves to a stationary state that defines the relationship between the state variables. From \eqref{error} we see that
$e_1$ and $e_3$ both go to zero asymptotically while $e_2$ tends to a constant value, which corresponds to the previously mentioned conserved quantity $K$. 
It should be mentioned here that a dynamical system made up of two coupled subsystems does not have independent variables upon complete synchronization and, hence, is overdetermined, since the variables of one subsystem are equal to the corresponding variables of the other subsystem. This is true for all synchronization systems that attain complete synchronization in the long term limit $t \to \infty$ \cite{pecora,kurths}. In the synchronization dynamics of systems that exhibit extreme multistability, the error dynamics also evolves to a stationary state, but now two of the corresponding variables of the subsystems are related by a conserved quantity, i.e., the variables are related by a constant (or a more complex relation) that depends on the initial conditions. This conserved quantity can be introduced into one of the subsystems as a bifurcation parameter that, while providing insights into the asymptotic behaviour, can be misinterpreted as a description of the overall synchronization dynamics. 
As discussed in
\cite{hens12}  the system (\ref{eqn}) possesses an infinite number of attractors corresponding to
different values of $K$. Further, the system admits
a Lyapunov function $V= e_1^2 + e_3^2$ such that $dV/dt < 0$ ensuring stability of the attractor states. 
As a numerical demonstration of the multiple attractor states we solve \eqref{eqn} for a fixed set of
system parameters ($\alpha= 0.02, a=0.2, b=0.2,c=5.7$) and vary the initial conditions $y_{20}$ from $-8.0$ to $7.0$ 
and $x_{20}$ from $-5.0$ to $0.0$
while keeping all 
the other initial conditions fixed at $x_{10}=y_{10}=z_{10}=z_{20} =0$. 
Fig.~\ref{fig:fig1}a  shows a plot of the maximum Lyapunov exponent
(color coded) in the ($x_{20} , y_{20}$) space. 
Fig.~\ref{fig:fig1}b shows the attractor states that exist along the line $x_{20}=-4.0$.

\begin{figure}[t!]
\includegraphics[width=8cm]{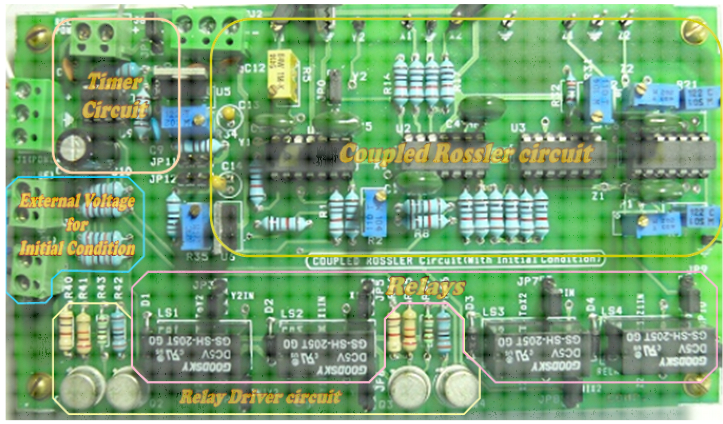}
\caption{(Color online) Experimental setup of the two coupled R\"{o}ssler oscillator system with relay and timer
circuits to change the initial conditions}
\label{device}
\end{figure}

We next turn to the experimental implementation of Eq.~\eqref{eqn}. Fig.~ \ref{device} shows a picture of the experimental set up that was constructed to 
study the dynamics of the coupled R\"{o}ssler system. A detailed circuit diagram of the system is available as supplemental material at \cite{supplement}.
A regulated power supply of $\pm 15V$ energizes the circuit, and the system parameters $\alpha$,
$a$, $b$ and $c$  are controlled with  circuit resistors. 
As a benchmark exercise, each individual R\"{o}ssler oscillator was separately tested by
varying the system parameters to obtain its various attractor states, ranging from periodic states
to chaotic dynamics, and the behavior was then compared to numerical simulations of the R\"{o}ssler
equations. Major care was taken to ensure that the two oscillator systems were as nearly
`identical' as possible within practical limits. This entailed careful weaning of all the component
elements (resistors, capacitors) to match their values as closely as possible and the removal of
any intrinsic drifts or biases within the operational amplifiers and multipliers \cite{circ}. The two oscillators 
were then coupled to each other as shown in Fig.~\ref{device} and their parameters fixed
at the values mentioned earlier. It should be mentioned that for the parameters chosen
in our experiment the individual R\"{o}ssler oscillators (when decoupled) were in the chaotic state.

\begin{figure*}
\raisebox{1.8cm} {(a)}\includegraphics[width = 0.15 \textwidth]{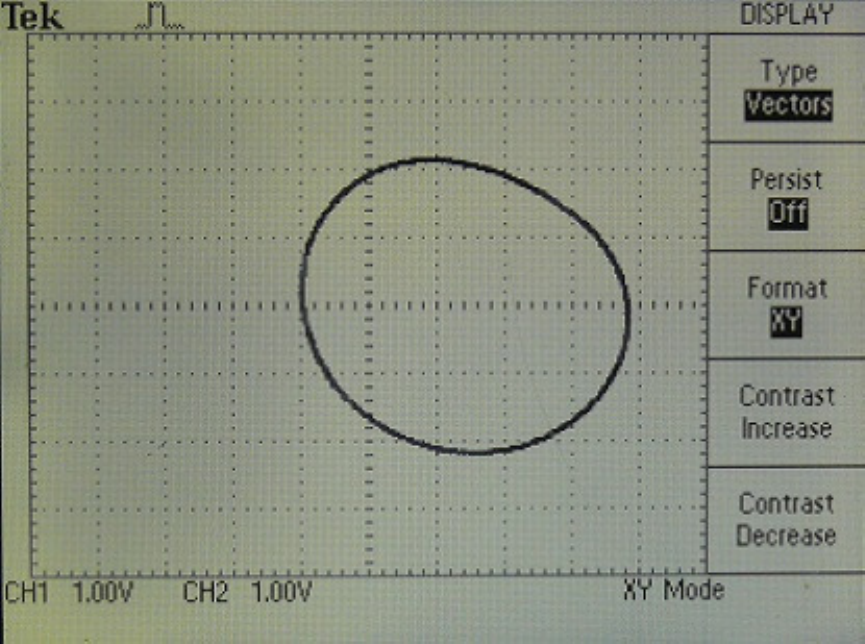}  
\raisebox{1.8cm} {(b)}\includegraphics[width = 0.15 \textwidth]{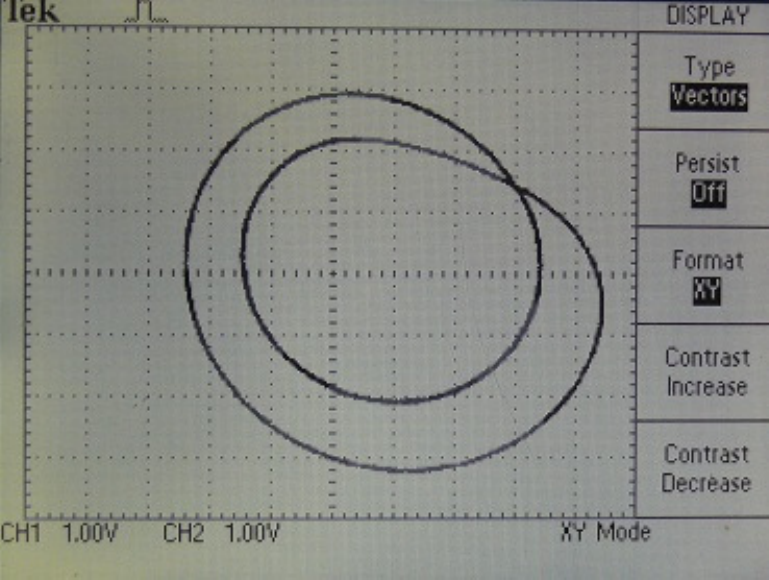} 
\raisebox{1.8cm} {(c)}\includegraphics[width = 0.15 \textwidth]{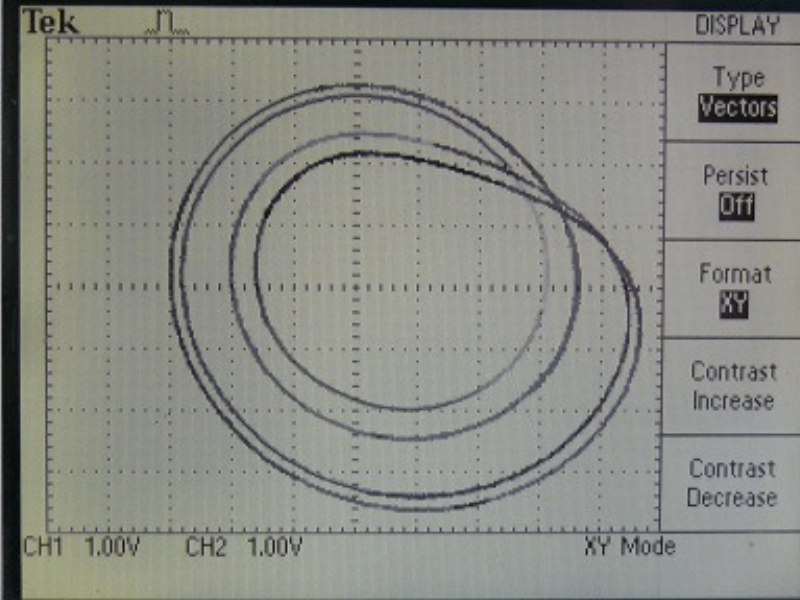} 
\raisebox{1.8cm} {(d)}\includegraphics[width = 0.15 \textwidth]{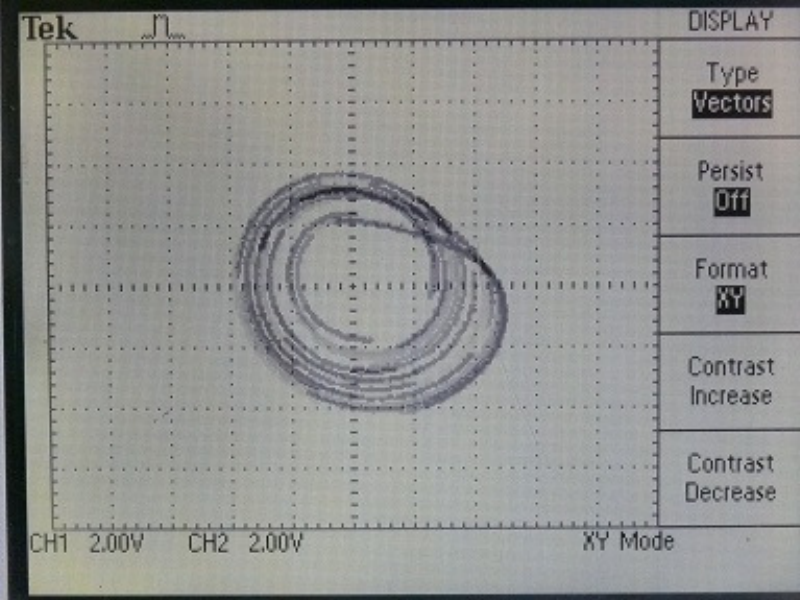}
\raisebox{1.8cm} {(e)}\includegraphics[width = 0.15 \textwidth]{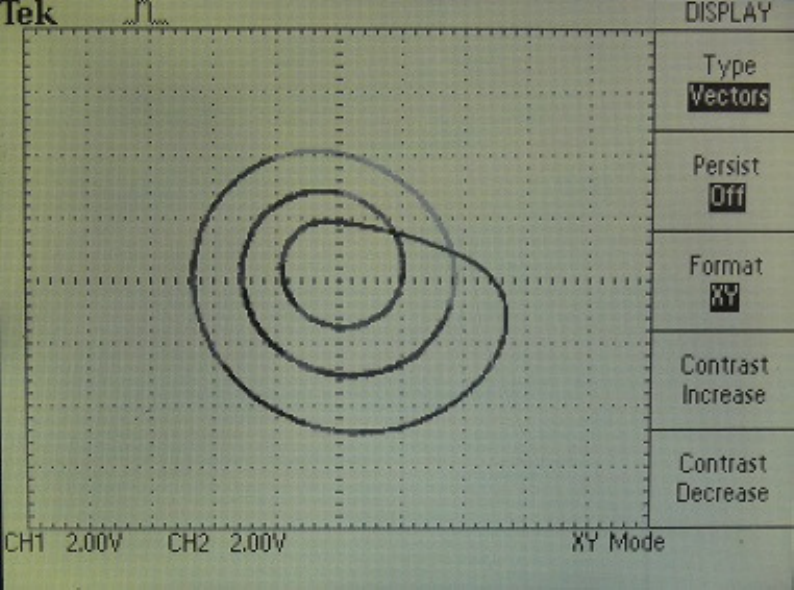} \\
\raisebox{1.8cm} {(f)}\includegraphics[width = 0.15 \textwidth]{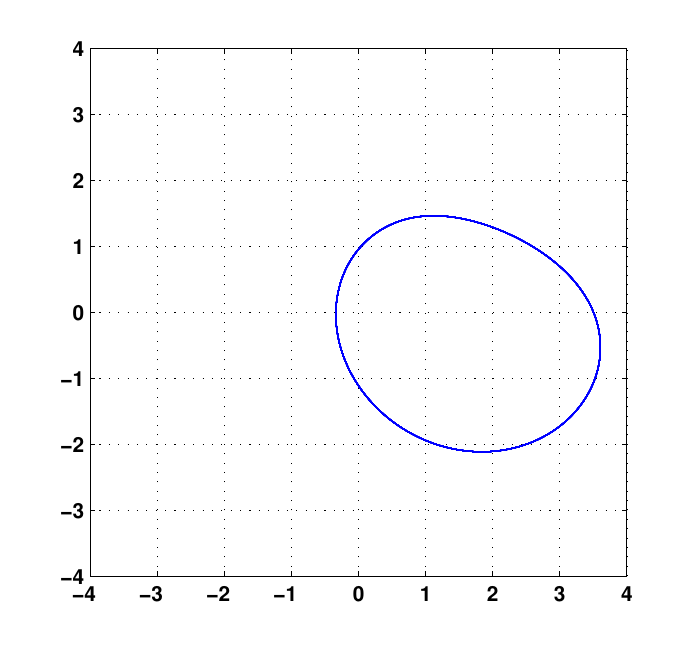} 
\raisebox{1.8cm} {(g)}\includegraphics[width = 0.15 \textwidth]{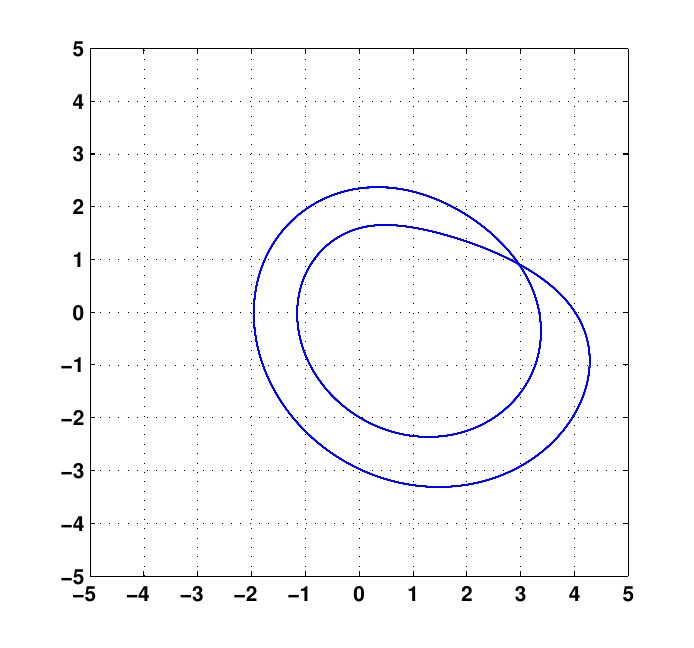}  
\raisebox{1.8cm} {(h)}\includegraphics[width = 0.15 \textwidth]{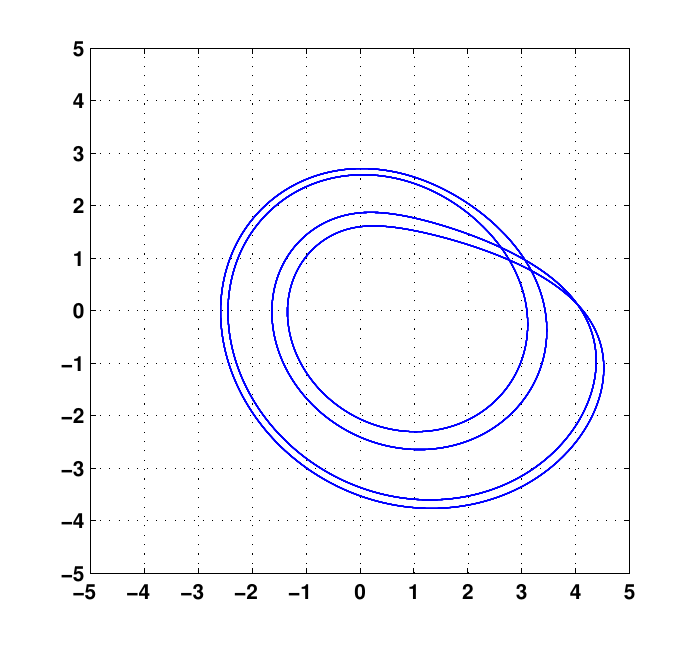} 
\raisebox{1.8cm} {(i)}\includegraphics[width = 0.15 \textwidth]{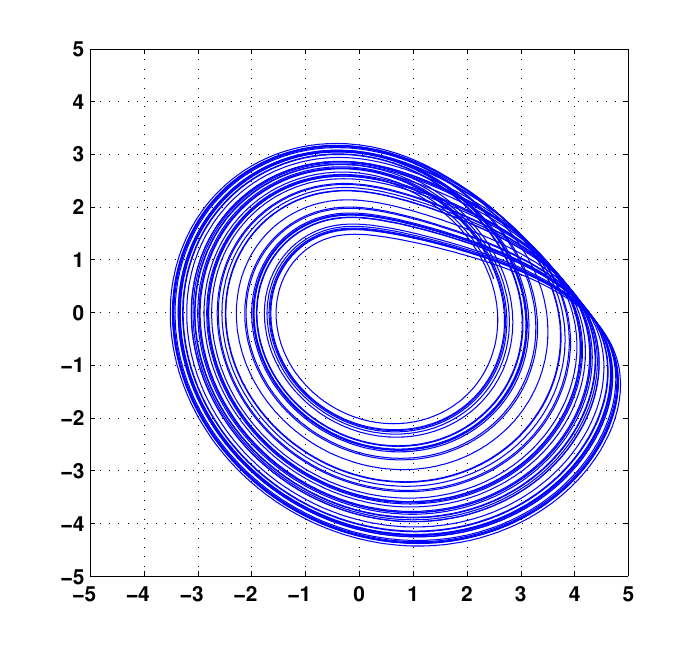} 
\raisebox{1.8cm} {(j)}\includegraphics[width = 0.15 \textwidth]{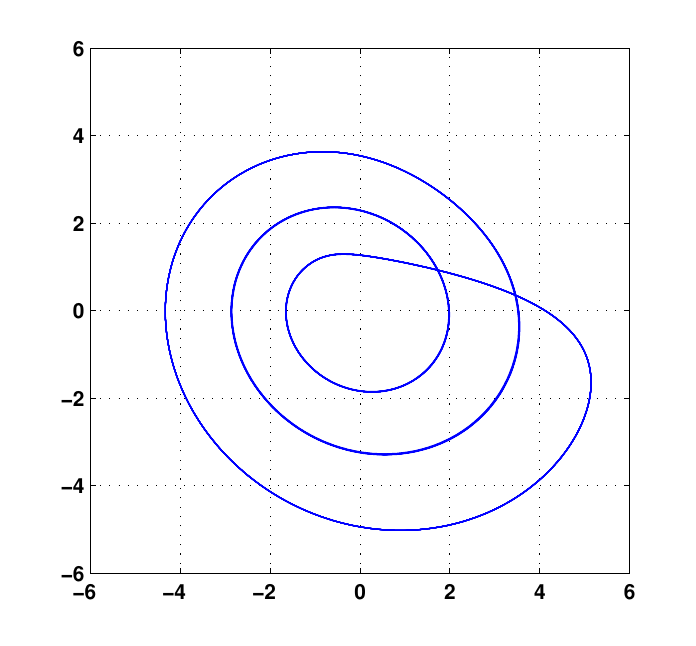}
\caption{(Color online) Oscilloscope images of the phase plots $y_1\; {\it vs}\; x_2$ of various attractor states with parameters $\alpha= 0.02, a=0.2, b=0.2,c=5.7$, 
initial conditions $x_{10}=y_{10}=z_{10}=z_{20} =0, x_{20}=-4.0$ and (a) $y_{20}=-7.0$, (b) $y_{20}=-4.0$, (c) $y_{20}=-2.0$, (d) $y_{20}=-1.0$, 
(e) $y_{20}=0.6$. The plots shown in (f), (g), (h), (i) and (j) are phase plots from numerical solutions of \eqref{eqn} corresponding to the initial conditions
of (a), (b), (c), (d) and (e) respectively.}
\label{phase_plots}
\end{figure*}

To change the initial conditions of the dynamics of the circuit, we have employed a strategy 
of imposing external voltages on selective nodes of the operational amplifiers as well as shorting
relevant capacitors of the circuit initially to set the values of some of the state variables to zero. 
To implement this combined strategy in a controlled manner we have developed and attached an additional system 
of relay circuits  to the coupled R\"{o}ssler circuits (indicated by a label in 
Fig. \ref{device}). For details of the circuit diagram of the relay circuit see \cite{supplement}.  
When this circuit is energized, the timer portion of the circuit produces a high output for $6$ seconds and the relay is turned on through the relay driver circuit. Due to this, the capacitor
responsible for generating the $x_{1}$ signal
gets shorted in the coupled circuit and hence $x_1$ is set to $0$V initially. Similarly $y_1, z_1$ and $z_2$ are also set to $0$V.  After $6$ seconds the output of the timer circuit falls to a low value and the relay gets switched off. Due to this, the capacitors no longer remain shorted  
and the circuit runs with the applied initial conditions for $x_1, y_1, z_1$ and $z_2$.  
The initial conditions of $x_2$ and $y_2$ are changed using independent
external voltage sources \cite{supplement}.
Using the above strategy we have run the coupled circuit for a number of initial conditions without changing the circuit parameters. 
Some typical results in the form of oscilloscope images of phase plots of $y_1\; {\it vs}\; x_2$ are shown in the top panel of Fig.~\ref{phase_plots} indicating the various periodic attractor states as well as a chaotic state. 


These states correspond to those identified
in the diagram of Fig. ~\ref{fig:fig1} by the labels of the images in Fig.~\ref{phase_plots}. 
The plots in the lower panel of Fig.~\ref{phase_plots} show corresponding numerical
solution results of Eq. (\ref{eqn}) using the same initial conditions.

One of the remarkable aspects of the synchronization dynamics that gives rise to extreme multistability, like in the coupled R\"{o}ssler circuit studied here, is that the coupled system has both dissipative dynamics and conservative dynamics. The dissipative dynamics is manifested in the nature of the dynamical state, which is a true attractor, with an infinite number of initial conditions that take the system to that attractor. The conservative dynamics is a consequence of the conserved quantity, which gives rise to a neutrally stable direction and, consequently, a dependence on the initial conditions. Thus the circuit is
characterized by infinitely many attractors, each associated with a particular value of the
conserved quantity $K = y_1 - y_2$ ($y_1$ and $y_2$ are asymptotic values taken at very large times), 
where the basin of attraction is made up of all sets of initial
conditions that evolve asymptotically to the particular value of $K$ associated with the
attractor. One can arbitrarily introduce a dependence on the initial conditions into any dynamical system described by a set of differential equations; however, in coupled systems undergoing synchronization, the conserved quantity arises from the synchronization dynamics.

The conserved quantity gives rise to a direction of neutral stability for the stationary states as well
as the periodic and chaotic orbits (in addition to that associated with the direction along the orbit). Hence, if the
system is in a particular periodic orbit, say period-2, a perturbation that does not satisfy the
condition of the conserved quantity will give rise to the evolution of the system to a new
attractor. The new attractor may differ only quantitatively; for example, a small perturbation
might shift the period-2 dynamics to a new period-2 dynamics that differs in amplitude.
However, larger perturbations give rise to the evolution of the system to qualitatively new attractors, such as a period-4 or
period-8 attractor. This characterization of the effects of perturbations also applies to the effects
of different initial conditions. Because the coupled system exhibits period-doubling bifurcations,
perturbations or different initial conditions permit the sampling of any of an infinite number of
qualitatively different attractors. Even if the coupled system did not display chaotic dynamics, the
same mechanism would give rise to an infinite number of quantitatively different attractors. 
In principle it should be possible to visit each of these asymptotic attractors in a continuous fashion by
altering the initial conditions in an infinitesimal manner - an extremely challenging task in an experimental
set-up.  However evidence of this continuous transition can be observed in a transient manner by deliberately introducing a slight
mismatch in the two circuits. We have carried out such an exercise (by slightly changing the value of one of
the resistors, see details in \cite{supplement}) such that the term $x_2- x_1$ on the RHS of eq.(\ref{b}) is changed to $x_2-x_1 +\delta x_2$ where $\delta$ is a small quantity. Then integrating
\eqref{eb}, one gets $e_2 = K - \alpha \delta \int{dt x_{2}(t)}$.   Thus the value of the constant $K$ that $e_2$ acquires changes with time and the rate
of change is controlled by the constants $\alpha$ and the mismatch value $\delta$. We have observed such a
continuous drift of the coupled system through the various states represented by Fig.~\ref{fig:fig1} when the 
two oscillators are slightly detuned by bringing about such a deliberate change of a small amount. The results can also
be reproduced exactly by a numerical solution of the mismatched system. The temporal
evolution through various states can be viewed in the video clip provided as supplementary material for
this paper \cite{supplement}. This demonstration provides additional support for
the existence of extreme multistability in the coupled R\"{o}ssler system \eqref{eqn}.\\

To conclude, we have described the first experimental demonstration of extreme multistability using an
electronic circuit implementation of two coupled R\"{o}ssler attractors. The theoretical model on which the experiment is based
has a simple set of equations. In particular, the error dynamics equations, \eqref{ea} - \eqref{ec}, of this model involve linear terms and are easily solvable. 
It is possible in principle to obtain extreme multistability from more complicated error dynamics that includes time dependent and nonlinear terms.  However
for the sake of simplicity and in the interest of minimizing the technical complexity of the experimental system we have avoided dealing with complicated error dynamics.\\
 
A restrictive feature of this type of dynamical system is the requirement that the coupled
subsystems be identical or nearly identical. This restriction as well as the need to devise a method to change the initial conditions in a controlled manner pose
serious technical challenges. We have successfully overcome these challenges in our electronic circuit system and demonstrated that
two chaotic circuits can be sufficiently matched to give rise to extreme multistability when appropriately coupled. It is likely that
extreme multistability will be a rarity in most physical, chemical and biological systems;
however, the combined conservative and dissipative features give rise to dynamics that might
find technological uses, such as the ability to easily select qualitatively different dynamical states
from an infinite number of possibilities. In addition, slightly mismatched systems that might occur in natural settings 
may display a temporal evolution through various dynamical states, as observed in our experimental electronic system.\\

{\it Acknowledgement:} U.F. would like to thank the Burgers Program for Fluid Dynamics of the University of Maryland for financial support. 

\bibliographystyle{apsrev}

\begin{thebibliography}{19}
\bibitem[{cir()}]{circ}
\bibinfo{note}{The multiplier used in the circuit is an {AD734} and the operational amplifiers are
  all {OP400}.}
\end{thebibliography}

\end{document}